\documentclass[twocolumn,showpacs,preprintnumbers,amsmath,amssymb,prl,aps,floatfix]{revtex4}

\usepackage{graphicx}
\usepackage{dcolumn}
\usepackage{bm}
\usepackage{natbib}

\begin{document}

\preprint{APS/PRB}

\title{Qubit Teleportation Using Doped Carbon Nanotubes}

\author{Abraham Alaka}
\altaffiliation[Also at ]{Electronics Department\\ University of York\\ Heslington York\\ YO10 5DD}
\email{chokosabe@gmail.com}
\affiliation{Electronics Department\\ University of York\\ Heslington York\\ YO10 5DD}%

\date{\today}

\begin{abstract}
We study theoretically, two doped Carbon nanotubes connected via a chemically active bond as the basic elementary gate for a recent proposal for qubit teleportation in the solid--state. We show that such a system provides the necessary entanglement between electron and electron--hole that is necessary for qubit teleportation. We simulate the transit of a defined qubit using time--dependent density functional theory within our structure. Finally, we critically discuss the implications of our design.  
\end{abstract}

\pacs{Valid PACS appear here}
\maketitle

\section{\label{sec:level1}Background}
For computers to become faster and faster, quantum computing must replace or supplement what we already have\cite{quantcomp}. To this end, there have been many proposals for structures capable of executing quantum computing. Proposed quantum computers utilise quantum degrees of freedom to create quantum bits (qubits). The intrinsic nature of the qubit allows for the performance of tasks such as factoring large integers and searching for and simulating quantum mechanical systems. However, the requirements of quantum computing present many new and unique challenges\cite{quantcomp}. Many of the current quantum computing proposals rely on the optical manipulation of information but they also have solid-state counterparts. 

A qubit is any system that can assume either of two super--posed values. It is normally composed of a two state system where the two states are simply called $\lvert0\rangle$ and $\lvert1\rangle$. The difference between a (classical) bit and a qubit being that the qubit lacks the characteristic (classical) energy barrier separating the states hence the qubit is neither fully $\lvert0$ or $\lvert1$ but a superposition of both.The qubit can be written as
\begin{equation}
\emph{qubit}=a_{0}\lvert0\rangle+a_{1}\lvert1\rangle
\end{equation}
where $\lvert0\rangle$ denotes the state in which the qubit has a value of 0 and $\lvert{1}\rangle$ denotes the state in whichthe qubit has a value of 1. The coefficients $a_{0}$ and $a_{1}$ are complex quantities whose squared magnitudes denote the probability that if a measurement is performed on the qubit, it will be found to have a value of 0 or 1 respectively.

A ready example of a qubit capable system is the spin--up or spin--down character of the electron. To date, many of the solid--state schemes proposed utilise the Nuclear spin. Naturally the spin of an electron would be the most natural candidate for a qubit but visualising a system capable of utilising the electron in this way often involves optical parallels. In which case the concept of a free electron is employed. To date, there is no such thing. 

Whilst qubit teleportation has been demonstrated to be possible for optically based schemes, there is still no directly viable occasion of qubit teleportation in the solid--state. The present day consensus is to move towards systems that can exhibit quantum computing characteristics and from there towards fully capable systems. In the solid--state case; this involves the demonstration of qubit teleportation.

A design for teleportation of the charge of an electron (and thus the qubit) within a "Fermi--sea" has been proposed by Prof. Beenakker and his group\cite{teleport}. A full summary is available from \cite{teleport}. In the following section we summarise their approach. In realising this scheme we confirm and expand upon the results of Gabor et al.\cite{gabor}. Our aim in this paper is to show how this scheme might be realised using current technology in the form of networks of connected carbon nanotubes doped with impurities. In realising quantum computation; the present challenge is to move towards systems where quantum computation phenomena can be demonstrated and controlled in experiment. It is with this view that we present this paper. 

\section{\label{sec:level2}Introduction}

\begin{figure}
\centering
\includegraphics[scale=0.7]{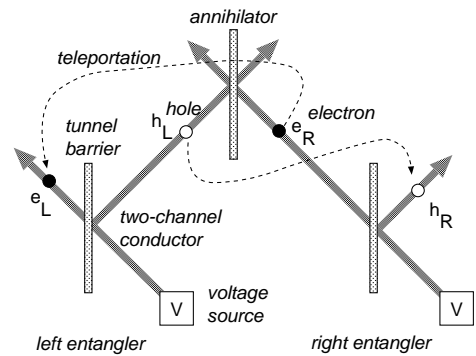}
\caption{Depiction of the teleportation scheme from \cite{teleport}. A voltage V is applied over a tunnel barrier produces pairs of entangled electron--hole pairs within the system. (e$_{L}$,h$_{L}$) are such a pair. As in \cite{teleport} we assume spin entanglement in the state ($\lvert\upuparrows\rangle + \lvert\downdownarrows\rangle)/\sqrt{2}$ where the first arrow refers to the electron spin and the second arrow refers to the hole spin. A second electron e$_{R}$ is in an unknown state $\alpha\lvert\uparrow\rangle+\beta\lvert\downarrow\rangle$. The electron e$_{R}$ can annihilate with the hole h$_{L}$ by tunnelling through the barrier at the centre. If it happens and is detected, then the state e$_{L}$ collapses to the state e$_{R}$. Note that $\lvert\uparrow\rangle$ annihilates with $\lvert\uparrow\rangle$ and $\lvert\downarrow\rangle$ annihilates with $\lvert\downarrow\rangle$ so e$_{L}$ inherits the coefficients $\alpha$ and $\beta$ of e$_{R}$ after its annihilation. The diagram shows a second entangler at the right, to perform two way teleportation. This essentially leads to entanglement swapping, e$_{L}$ and h$_{R}$ become entangled after the annihilation of h$_{L}$ and e$_{R}$}
\label{scheme}
\end{figure}

Electron tunnelling is a manifestation of the Heisenberg Uncertainty principle and quantum mechanics. An electron hole is an unoccupied state within a Fermi--sea. It is one of only two charge carriers in the solid--state\footnote{The electron being the second.} The opposing charges of hole and electron means that they can only exist with a separating insulating barrier between them. If this barrier passes a small current and an electron meets a hole, then both are annihilated. The proposed scheme works with the use of electron and hole pairs separated by potential barriers, Figure \ref{scheme} is a simple schematic.

At low temperatures, all energy levels are filled with electrons to the maximum energy with the maximum energy being the Fermi--energy. This is referred to as the Fermi sea. This 'sea' is in equilibrium and thus carries no current (excitations may induce a current). The tunnelling event happens when the Fermi energy on one side of the barrier is higher than that on the other side of the barrier. So the hole is raised to the same energy as the electron. There is a probability that the electron and the hole are reflected by the barrier (depending on the conditions) but each time they meet, there is also a probability that the electron will tunnel through the barrier and fall into the hole on the other side. This mechanism forms the basis of the teleport proposal. 

In its simplest form, the teleportation process begins with the entangled state $(\lvert\uparrow\rangle_{e}\lvert\uparrow\rangle_{k})/\sqrt{2}$\cite{proposal}. The subscripts e and h refer to the electron and the hole at two different locations respectively. The particle teleported is another electron in the state $\alpha\lvert\uparrow\rangle_{\acute{e}}+\beta\lvert\downarrow\rangle_{\acute{e}}$ with $(\lvert\alpha\rvert^{2} +\lvert\beta\rvert^{2}=1)$. The second electron $\acute{e}$ can tunnel into the hole h only if the spins match. The probability of this happening can be expressed as $\frac{1}{2}\lvert t\rvert^{2}=\frac{1}{2}\lvert\alpha\rvert^{2}\lvert t\rvert^{2}+\frac{1}{2}\lvert\beta\rvert^{2}\lvert t\rvert^{2}\ll1$ where $t$ is the tunnelling amplitude. There is no instantaneous transfer of information. The unpredictability of the success of the tunnelling attempt means that a message will need to be sent by classical means that the teleportation has happened.

The instantaneous transfer of the state from one to the other electron is necessary to satisfy the no--cloning theorem of quantum mechanics\cite{quantcomp}. At no point is there more than one copy of the state. As usual, all the limitations of teleportation still apply\cite{quantcomp,proposal}. We should note that whilst qubit teleportation has been demonstrated to be possible for optically based schemes, there is still no directly viable occasion of qubit teleportation in the solid--state. 

\begin{figure}
\centering
\includegraphics[scale=0.5]{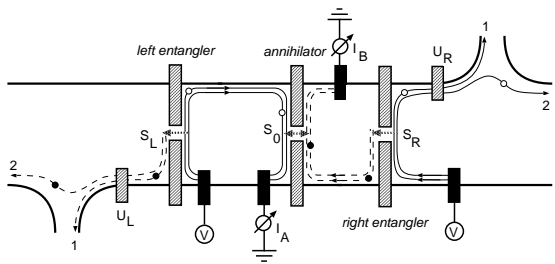}
\label{pe}
\caption{Schematic description of a method to produce and detect entangled edge channels in the quantum Hall effect from \cite{proposal}. The bold black lines mark the boundaries of a two--dimensional electron gas. A strong perpendicular magnetic field B ensures that the transport near the Fermi--level $E_{F}$ takes place in two edge channels, extended along a pair of equipotentials (thin solid and dashed lines, with arrows that give the direction of propagation). The split gate electrode (shaded rectangles at the center) divides the conductor into two halves, coupled by tunnelling through a narrow opening (dashed arrow). If a voltage is applied between the two halves, then there is a narrow energy range $eV$ above $E_{F}$ in which the edge channels are predominantly filled in the left half (solid lines). Tunnelling events introduce filled states in the right--half (black dots) and empty states in the left half (open circles). The entanglement of these particle--hole excitations is detected via the violation of a Bell inequality.}
\end{figure}

Central to the design of this 'qubit--teleporter' is the architecture that allows for the production (and detection) of entangled electron--hole pairs in the Fermi sea. Figure \ref{pe} (from the same group) shows how this might be achieved. This would be the building block of any solid--state solution for the proposed scheme.

\section{The System}
Our aim in this paper is to highlight a system capable of this entanglement (and ultimately qubit teleportation) in the form of impurity doped carbon nanotubes connected via chemically active bonds. Previous studies\cite{gabor,msc} have shown that suitably doped semi--conducting carbon nanotubes form chemically active C--C bonds when placed in proximity to each other. These bonds can serve as the necessary tunnel barriers depicted in Fig \ref{scheme}; thus allowing for its realisation. 

Previous studies have also indicated that semiconducting carbon nanotubes with chirality (8,0) are well suited to such a role. LDA band structure calculations predict a minimum band gap of 0.6 eV whilst Quasiparticle corrections open this up to 1.75eV\cite{beigi}. Quasiparticle  band structure calculations have shown nanotubes with chirality (8,0) to be of an LDA determined 1.75 eV with a stable structure diameter of < 1nm.
 
The nodes in the circumferential direction within this structure are also quantised. With the inclusion of impurities (via doping), carbon nanotubes exhibit a reduced band gap and excess spin--density localisation\cite{gabor,msc} localised about the impurity. The studies of \cite{beigi} have also indicated that the confinement of electrons and holes within the un--doped structure is of the order of several \AA. Within our planned structure this has the effect of increasing the probability of electron tunnelling and entanglement swapping.

We simulate 2 doped semiconducting carbon nanotubes connected via a chemically active bond as our 'quantum entangler'. The tubes are of chirality (8,0) and doped with Nitrogen (N). The tunnel barrier between the sections of nanotube is analogous to the split gate electrode in Fig \ref{pe}. The application of a voltage across either of the tubes (across the tunnel barrier) then results in tunnelling of an electron on one side to an empty state on the other side. In our calculations we simulate 2 sections both 2 unit cells long in the Z direction and each with a nitrogen impurity.  

Regard to the information contained on the tunnelling electron means that the barrier must be shielded from interference for its surrounding environment. A recent study of super--luminal electron tunnelling suggests that the information of a tunnelling electron wave--packet is essentially readjusted within the barrier\cite{tunnel1,tunnel2}. In effect, the dwell time of the electron \footnote{If it does exist} must be kept to a minimum to ensure the integrity of the wavepacket (which is the spin character of the electron). Furthermore, to ensure entanglement, the material should be a semi--conductor to preserve the initial state of the qubit and allow for the needed decoherence times. Our choice of the (8,0) carbon nanotube meets all these criteria.
 
Our design exploits the widely reported and verified properties of carbon nanotubes. Within the configuration laid out above, the electron and hole are both confined to a 1 dimensional line. The only degree of freedom allowed is through the tunnel barrier connecting the tubes. Any other attempt to escape the structure incurs a massive energetic penalty. In the case of carbon nanotubes (with a band gap of < 2ev) the potential is still enough to confine the electron and hole even at room temperatures. In the case of a defined qubit (which is weakly coupled to its environment), the spatial degree of freedom allowed by the tunnel barrier heightens the tunnelling probability. 

\begin{figure}
\centering
\includegraphics[scale=0.3]{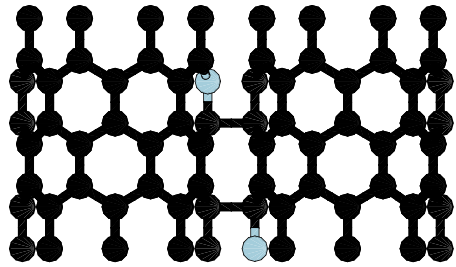}
\caption{Classical depiction of two sections of semiconducting Carbon nanotubes connected by chemically active bonds. Both sections of tube are of chirality (8,0) and 2 unit cells in length. The blue spheres denote impurity sites (Impurities being Nitrogen atoms). The excess electrons that the impurities carry are effectively localised within the structure\cite{msc,gabor}.}
\label{tubes}
\end{figure}

The use of semiconducting carbon nanotubes in particular offers many advantages. Among these are the fact that advances in fabrication are making the production of carbon nanotubes less and less expensive. There is also the fact that carbon nanotubes are very good structures for conserving the spin of the charge carrier, thus preserving the integrity of the qubit over the length scales involved. Another advantage is that one then has the option of layout and circuit design and all this is within the reach of current technology. An illustration of our proposed design is Fig \ref{tubes} 

Figure \ref{tubes}.

\section{\label{sec:level3}Method}
Within our simulation cell, we define of two doped sections of carbon nanotube with chirality (8,0) and composed of 2 primitive unit cells (each primitive unit cell contains 32 atoms) in the z direction. The atomic coordinates are generated using the program Tubegen\cite{tubegen} with a C--C bond length of 1.425\AA (as confirmed by experiments). Each section has a single nitrogen impurity in place of a C atom and both are in the x--plane facing each other in the middle as shown in Figure \ref{tubes}. There is an inter--tube distance of 1.6\AA (half the interlayer distance in natural graphite).

We define an excess spin density in the form of an electron with a spin--up character. We assign this excess spin density to the tube section to the left in Figure\ref{tubes} as indicated in Figure\ref{plot}. Due to computational constraints, we only simulate 2 sections of carbon nanotube as opposed to infinite length tubes. However, with our approach, our results are easily applicable to the infinite length tube case. To account for our finite system we perform total energy calculations using the Louis Bearends (LB) functional\cite{lb94}. The LB functional is of a form designed to get the correct asymptotic behaviour; yielding good energies and excitation energies in calculations of response properties.

We use improved troullier--martins pseudopotentials within the GGA approximation as implemented in the octopus code\cite{octopus}. The pseudopotentials are distributed with the package and they have been extensively tried and tested in many different applications. We perform static total--energy calculations and find that when these localised impurities face each other they form a chemically active bond connecting the two sections of carbon nanotube together (confirming the results of \cite{gabor}.

We then compute the time--dependent simulation of a polarised electric field across the tubes. The electric field is of delta strength (voltage) $0.05$eV${-1}$ polarised -100\% in the x direction and 0\% in the y and z directions respectively. We propagate the time--dependent Kohn--Sham equations using the magnus evolution operator\cite{magnus}. at a time step of 0.0005ev for a total of 2000 time steps. From a multipole representation of the time propagation, we calculate the evolution of the dipole--strength function of the system. In experiment, the optical--absorption cross section is directly accessible but in this instance, our use of the dipole strength function is of a qualitative nature. It presents a ready picture of the shifts in polarizability across the system as well as serving as a means of identification of such a system.


Our use of the dipole strength function serves as a poor man's measure of the electron hole interaction. Previous studies \cite{beigi} have shown that the optical response of (8,0) carbon nanotubes is dominated by excitonic terms.  Contained within the qualitative expression of the dipole strength function are the excitonic effects between the electron, the hole and the hole created by the transition of the electron across the tunnel barrier. The changing values (and sign) of the dipole strength function with respect to energy is indicative of the electron--hole interaction. 

To calculate the dipole--strength function (which is simply proportional to the absorption cross section), we first calculate the ground--state energy and then excite the system with an (afore--mentioned) delta electric field, $E_{0}\delta(t)$. The dipole strength can then be related to to the imaginary part of the dynamical polarizability by
\begin{equation}
S(\omega)=\frac{4\pi m_{e}}{h^{2}}\omega\mathcal{J}\alpha(\omega)
\end{equation}
where $h$ is Planck's constant, $m_{e}$ the electron mass and the dynamical polarizability,
\begin{equation}
\alpha(\omega)=-\frac{2}{E_{0}}\int dr\mathcal{Z} \delta n(\bold{r},\omega).
\end{equation}
In the last expression, $\delta n(\bold{r},\omega)$ stands for the Fourier transform of $n(r,t)-n(r,t=0)$. We present our results below.

\section{Results}
\begin{figure}
\centering
\includegraphics[scale=0.6]{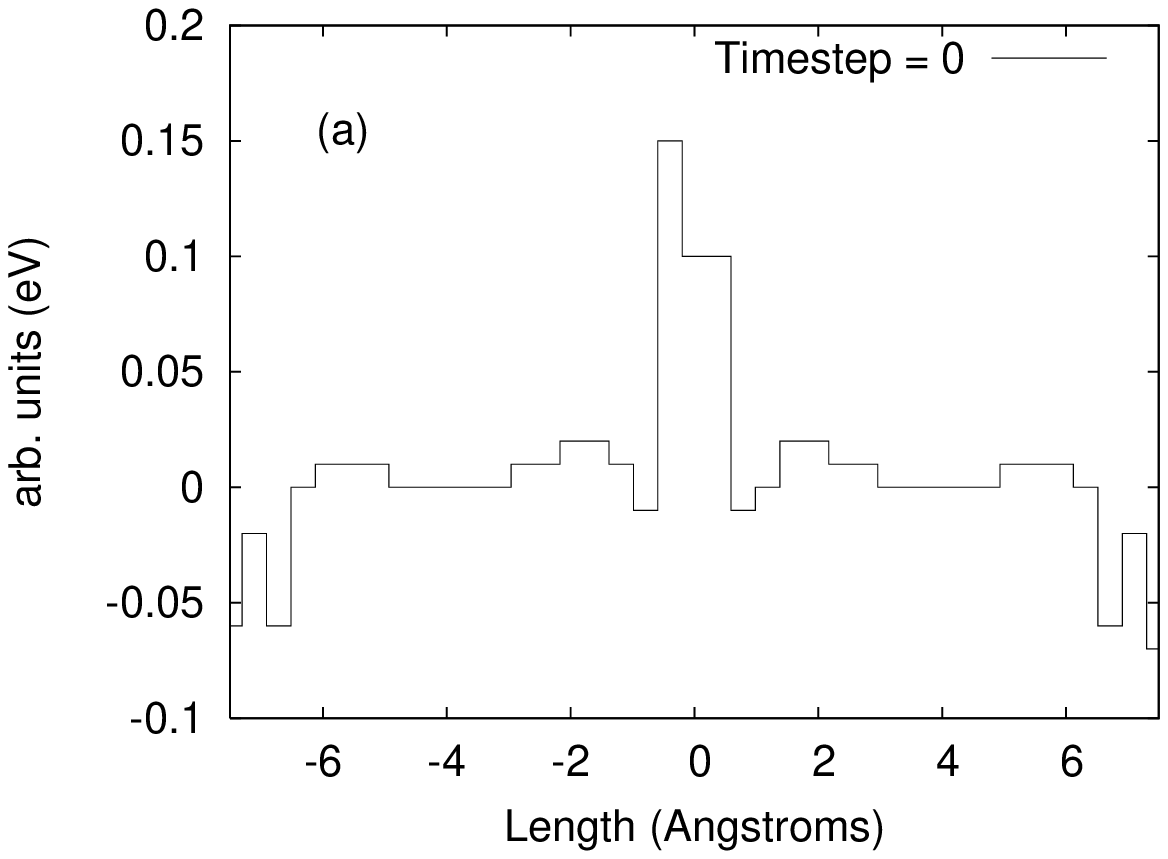}
\includegraphics[scale=0.6]{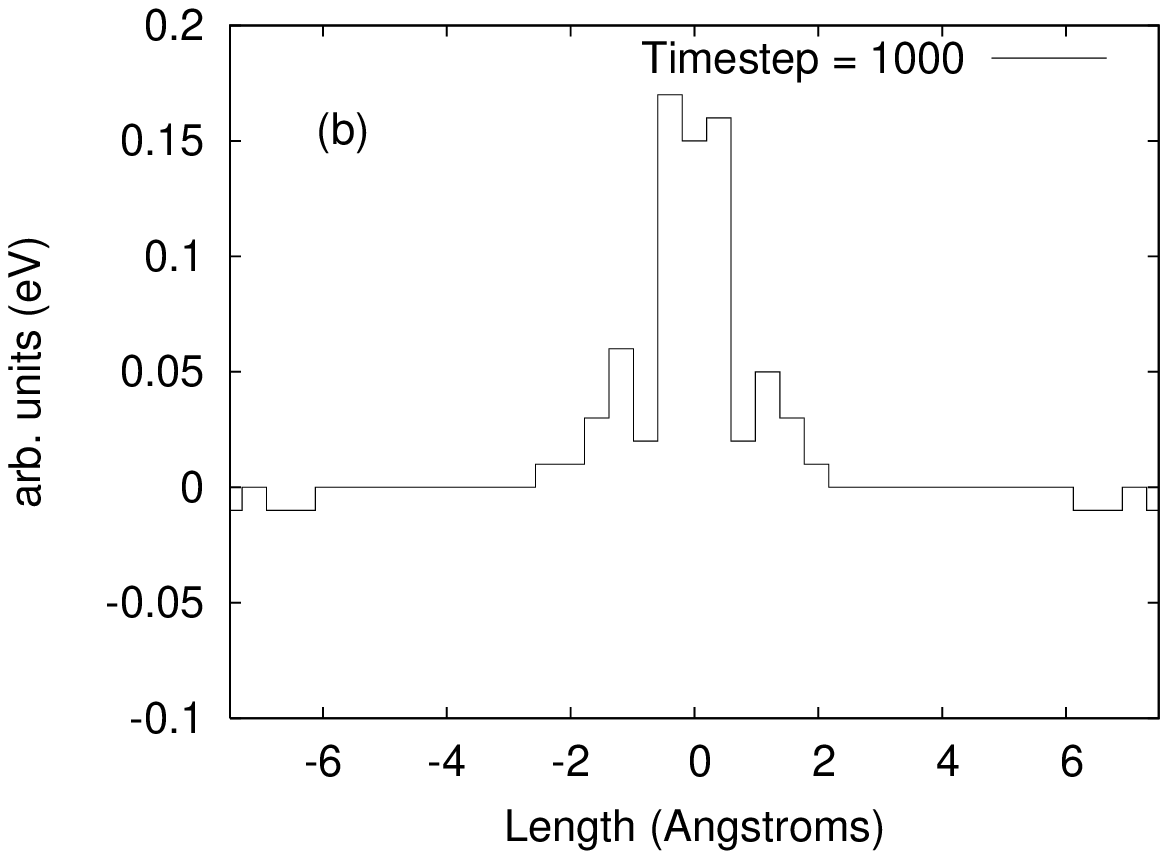}
\includegraphics[scale=0.6]{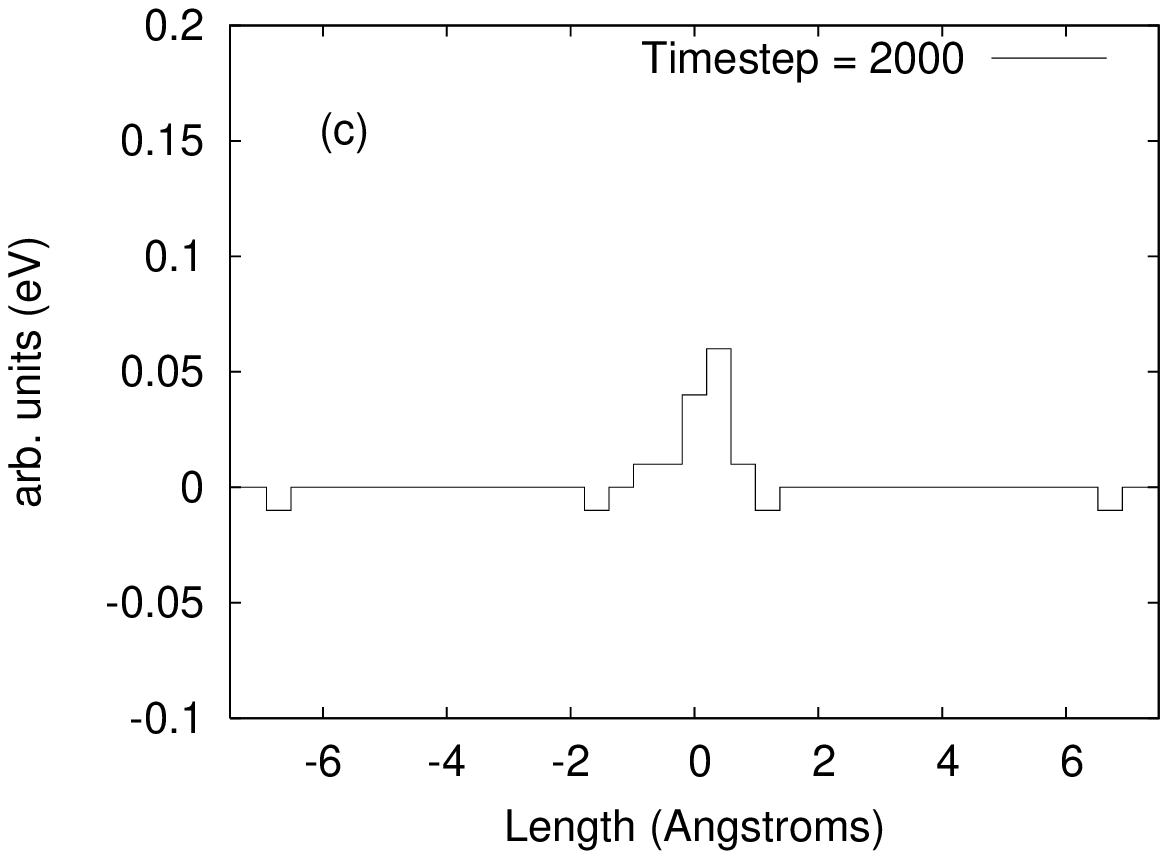}
\caption{The Un--paired Spin--density of the system projected onto the x--axis. In (a) we see a clearly defined qubit. (b) shows the propagation of this qubit across the tunnel barrier. (c) represents the position of the qubit post time propagation. Throughout it can be seen that the opposing and resultant electron--hole is delocalised across the structure. The electron is localised throughout.}
\label{plot}
\end{figure}

\begin{figure}
\centering
\includegraphics[scale=0.2,angle=-90]{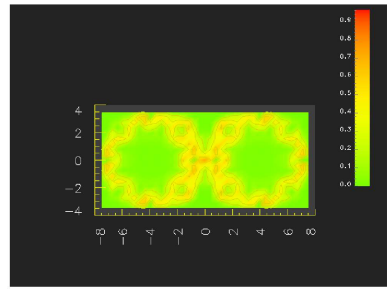}
\label{elf}
\caption{A cross section of the electron localisation function of the system from the static total energy calculation projected onto the z--axis. Green areas indicate areas of lowest localisation whilst red areas indicate the maximum. There is a maximum ELF value across the bond of 0.9.}
\end{figure}

From the initial static total energy calculation of the system, we find that the maximal force on any atom in the simulation cell to be less than 150(eV/A). This may be irrelevant given the inaccuracy of the LB functional in calculating forces but the figure is still within a magnitude of what is acceptable for our design. The direct substitutionality of the impurity atoms is confirmed; the maximum forces acting on the impurity atoms are less than 10(eV/A). We also find that at an intertube distance of 3.2 \AA, (the distance between graphite layers in nature) the doped sections of tube form one main intertube C--C bond as opposed to the classical picture shown in Figure \ref{tubes}. A cross section of the electron localisation function (ELF) of the system is shown in Figure \ref{elf}. It can be seen that about midpoint along the length (in the z--direction) of both tubes, there is a maximum electron localisation value ~ 0.8 (representing the bond) and nominal values of 0.5 on either side of the bond representing spin--density accumulation. Analysis of the final Spin--Density configuration of the system shows that the applied electronic excitation in the form of the applied E-field propels this excess spin density across the bond. This is mapped out in the plots of the Un--paired spin--density (Figures \ref{plot}(a),(b) and (c)) at different timesteps.

From a static representation of our system (with a defined excess qubit) we generate a picture where we have a qubit value (a clearly defined spin--up character to the spin--density and an ELF function of the chemical bond in Fig \ref{plot}(a). The effect of the applied field is to propel this excess spin density from the left side of the system across the bond to the right side of the system as shown in Figure \ref{plot}(c). Analysis of the dipole strength function shows shifts in the orientation of the dynamic polarizability of the system. This is in part due to the geometric shifts in the position of our defined qubit across the system. 

\section{Discussion}

The bond formed between the tubes is covalent and thus not easily broken. With an ELF value of 0.9, the bond is prevalent even at temperatures elevated beyond the ground state.  We should note that the above example is merely illustrative of what is possible, this same scheme as applied to infinite length tubes in networks formed by chemically active bonds favours the teleportation proposal more so. In the case of infinite length tubes there would be even more of a localisation and isolation of the qubit wave--packet\cite{gabor,msc}. We note that in real applications, the tunnelling probability of the qubit would be very low. However, the system described above is easily scaleable to suit the design depicted in Figure \ref{scheme}. The optical properties of nanotubes with chirality of (8,0) is strongly dominated by excitonic effects. This feature also bodes much for our proposal. Within the setup outlined above the availability of excitons on either side of the tunnel barrier greatly increases he probability of a tunnelling event and hence entanglement swapping. However, we should also note the the effects of electron phonon coupling also have a part to play.

One of the said requirements of a  system capable of quantum computing is one which is easily scalable with well characterised qubits. We believe our design to be a ready example of such a system. In implementation, this allows for the expression of the spin 1/2 value of the electron as 
\begin{equation}
\mid\uparrow\rangle\equiv\mid1\rangle
\end{equation} 
 
In an experimentally viable setup, we expect that any device would be of the form of Fig \ref{bulk}
An arrangement such a this would have a shear modulus of ~80 Gpa or more\cite{gabor}.   

As concerns a traversal time or a tunnelling time, a true definition is (not or might not ever) be possible\cite{tunnel1} (an un--controversial definition still escapes the literature.) However, within the context of this work, the 'view--point' of the qubit is what we are interested in and what we 'impose' on the simulation. True electron tunnelling (without distortion) is of itself a \emph{quasistatic} process in which the output density (from the tunnel barrier) adiabatically follows the input with a small delay due to energy storage within the barrier\cite{tunnel2}. This time--delay is (probably) the most representative of what we might define as an operation time for the device, or a dwell time for the information carrier. In this simulation, this time was of the order of 2eV$^{-1}$ (at an intertube distance of 1.6\AA). 

In general, the scheme as approximated in this model can only be a guide as to the decoherence times one might expect of the defined qubit. As a rough guide, we employ the equation of Burkard and Loss; in which they posit that the spin coherence time may be measured as a function of the Electron Spin Resonance (ESR). The resulting stationary current would exhibit a resonance whose line width is determined by this time. The ultimate feasibility of such a configuration must be extracted from experiment; our results clearly support such a configuration. 

\begin{figure}
\centering
\includegraphics[scale=0.6]{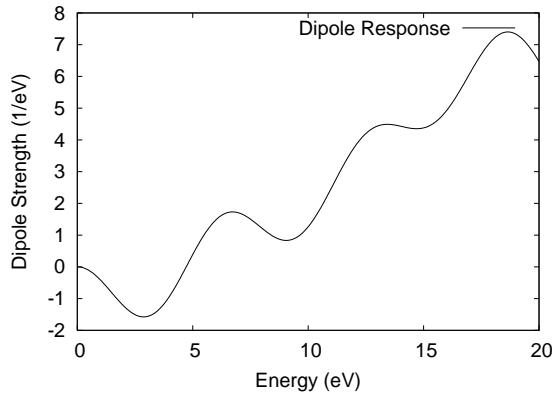}
\label{dips}
\caption{The dipole strength function of the system over a time--interval representative of the qubit transition time. A Comparison with Energy indicates multiple peaks and troughs. The first of which we define as the deterministic point of tunnelling. This generated optical signature may serve as an aid to the detection of the event within a bundle of nanotubes.}
\end{figure}

\section{Conclusions}
In conclusion, We have performed total--energy pseudopotential calculations on two sections of doped carbon nanotube using the plane--wave pseudopotential code octopus\cite{octopus}. We have shown that when the impurities face each other, a bond forms between the two sections of carbon nanotube (confirming the results of \cite{gabor}. We have shown that this chemically active bond may be used as a tunnel barrier in the realisation of the solid--state entanglement scheme of \cite{teleport}. Finally, we have shown that such a system allows for electron hole entanglement that makes the solid state teleportation scheme depicted in \cite{teleport} work. Extending this, we have identified an optical profile in the form of a qualitative dipole response which may be used to identify and isolate such a device within a bundle of tubes.

\newpage 


\begin{thebibliography}{12}
\expandafter\ifx\csname natexlab\endcsname\relax\def\natexlab#1{#1}\fi
\expandafter\ifx\csname bibnamefont\endcsname\relax
  \def\bibnamefont#1{#1}\fi
\expandafter\ifx\csname bibfnamefont\endcsname\relax
  \def\bibfnamefont#1{#1}\fi
\expandafter\ifx\csname citenamefont\endcsname\relax
  \def\citenamefont#1{#1}\fi
\expandafter\ifx\csname url\endcsname\relax
  \def\url#1{\texttt{#1}}\fi
\expandafter\ifx\csname urlprefix\endcsname\relax\def\urlprefix{URL }\fi
\providecommand{\bibinfo}[2]{#2}
\providecommand{\eprint}[2][]{\url{#2}}

\bibitem[{\citenamefont{Ekert and Jozsa}(96)}]{quantcomp}
\bibinfo{author}{\bibfnamefont{A.}~\bibnamefont{Ekert}} \bibnamefont{and}
  \bibinfo{author}{\bibfnamefont{R.}~\bibnamefont{Jozsa}},
  \bibinfo{journal}{Reviews of Modern Physics} \textbf{\bibinfo{volume}{68}},
  \bibinfo{pages}{733} (\bibinfo{year}{96}).

\bibitem[{\citenamefont{Beenakker and Kindermann}(2003)}]{teleport}
\bibinfo{author}{\bibfnamefont{C.~W.~J.} \bibnamefont{Beenakker}}
  \bibnamefont{and}
  \bibinfo{author}{\bibfnamefont{M.}~\bibnamefont{Kindermann}},
  \bibinfo{journal}{Physical Review Letters} \textbf{\bibinfo{volume}{92}},
  \bibinfo{pages}{056801} (\bibinfo{year}{2003}).

\bibitem[{\citenamefont{Nevidomsky et~al.}(2003)\citenamefont{Nevidomsky,
  Csanyi, , and Payne}}]{gabor}
\bibinfo{author}{\bibfnamefont{A.~H.} \bibnamefont{Nevidomsky}},
  \bibinfo{author}{\bibfnamefont{G.}~\bibnamefont{Csanyi}}, , \bibnamefont{and}
  \bibinfo{author}{\bibfnamefont{M.~C.} \bibnamefont{Payne}},
  \bibinfo{journal}{Physical Review Letters}  (\bibinfo{year}{2003}).

\bibitem[{\citenamefont{Beenakker et~al.}(2003)\citenamefont{Beenakker,
  C.~Emary, and van Velsen}}]{proposal}
\bibinfo{author}{\bibfnamefont{C.~W.~J.} \bibnamefont{Beenakker}},
  \bibinfo{author}{\bibfnamefont{M.~K.} \bibnamefont{C.~Emary}},
  \bibnamefont{and} \bibinfo{author}{\bibfnamefont{J.~L.} \bibnamefont{van
  Velsen}}, \bibinfo{journal}{Physical Review Letters}
  \textbf{\bibinfo{volume}{91}}, \bibinfo{pages}{147901}
  (\bibinfo{year}{2003}).

\bibitem[{\citenamefont{Alaka}(2004)}]{msc}
\bibinfo{author}{\bibfnamefont{A.}~\bibnamefont{Alaka}}, Master's thesis,
  \bibinfo{school}{University of York} (\bibinfo{year}{2004}).

\bibitem[{\citenamefont{Spataru et~al.}(2004)\citenamefont{Spataru, Beigi,
  Benedict, and Louie}}]{beigi}
\bibinfo{author}{\bibfnamefont{C.~D.} \bibnamefont{Spataru}},
  \bibinfo{author}{\bibfnamefont{S.~I.} \bibnamefont{Beigi}},
  \bibinfo{author}{\bibfnamefont{L.~X.} \bibnamefont{Benedict}},
  \bibnamefont{and} \bibinfo{author}{\bibfnamefont{S.~G.} \bibnamefont{Louie}},
  \bibinfo{journal}{Physical Review Letters} \textbf{\bibinfo{volume}{92}},
  \bibinfo{pages}{077402} (\bibinfo{year}{2004}).

\bibitem[{\citenamefont{Landauer and Martin}(1994)}]{tunnel1}
\bibinfo{author}{\bibfnamefont{R.}~\bibnamefont{Landauer}} \bibnamefont{and}
  \bibinfo{author}{\bibfnamefont{T.}~\bibnamefont{Martin}},
  \bibinfo{journal}{Reviews of Modern Physics} \textbf{\bibinfo{volume}{66}},
  \bibinfo{pages}{217} (\bibinfo{year}{1994}).

\bibitem[{\citenamefont{Winful}(2003)}]{tunnel2}
\bibinfo{author}{\bibfnamefont{H.~G.} \bibnamefont{Winful}},
  \bibinfo{journal}{Physical Review Letters} \textbf{\bibinfo{volume}{90}},
  \bibinfo{pages}{023901} (\bibinfo{year}{2003}).

\bibitem[{TubeGen 3.1, J. T. Frey and D. J. Doren, University of Delaware,
  Newark DE, 2003()}]{tubegen}
TubeGen 3.1, J. T. Frey and D. J. Doren, University of Delaware, Newark DE,
  2003.

\bibitem[{\citenamefont{Leeuwen and Baerends}(1994)}]{lb94}
\bibinfo{author}{\bibfnamefont{R.~V.} \bibnamefont{Leeuwen}} \bibnamefont{and}
  \bibinfo{author}{\bibfnamefont{E.~J.} \bibnamefont{Baerends}},
  \bibinfo{journal}{Physical Review A} \textbf{\bibinfo{volume}{49}},
  \bibinfo{pages}{2421} (\bibinfo{year}{1994}).

\bibitem[{\citenamefont{Marques et~al.}(2003)\citenamefont{Marques, Castro,
  Bertsch, and Rubio}}]{octopus}
\bibinfo{author}{\bibfnamefont{M.~A.~L.} \bibnamefont{Marques}},
  \bibinfo{author}{\bibfnamefont{A.}~\bibnamefont{Castro}},
  \bibinfo{author}{\bibfnamefont{G.~F.} \bibnamefont{Bertsch}},
  \bibnamefont{and} \bibinfo{author}{\bibfnamefont{A.}~\bibnamefont{Rubio}},
  \bibinfo{journal}{Computational Physics Communications}
  \textbf{\bibinfo{volume}{151}} (\bibinfo{year}{2003}).

\bibitem[{\citenamefont{Hochbruck and Lubich}(2003)}]{magnus}
\bibinfo{author}{\bibfnamefont{M.}~\bibnamefont{Hochbruck}} \bibnamefont{and}
  \bibinfo{author}{\bibfnamefont{C.}~\bibnamefont{Lubich}},
  \bibinfo{journal}{SIAM Journal of Numerical Analysis}
  \textbf{\bibinfo{volume}{41}}, \bibinfo{pages}{245} (\bibinfo{year}{2003}).

\end{thebibliography}
\end{document}